\documentclass[manuscript,nonacm]{acmart}

\AtBeginDocument{%
  }

\setcopyright{acmlicensed}
\copyrightyear{2024}
\acmYear{2024}
\acmDOI{XXXXXXX.XXXXXXX}

\acmConference[]{}{}{}
\acmISBN{978-1-4503-XXXX-X/18/06}

\usepackage{colortbl}
\usepackage{hhline}
\usepackage{xspace}
\usepackage{color}
\usepackage{placeins}
\usepackage{booktabs}

\usepackage{pgfplots}
\pgfplotsset{compat=1.9}

\definecolor{highpp}{RGB}{51,34,136}
\definecolor{overall}{RGB}{17,119,51}
\definecolor{lowpp}{RGB}{136,204,238}

\definecolor{colortwo}{RGB}{149, 165, 166} 
\definecolor{colorthree}{RGB}{46, 204, 113}

\newcommand{\etal}{et al.\xspace}
\newcommand*{\eg}{e.g.,\xspace}

\newcommand{\oneonly}{\textbf{P1}\xspace}
\newcommand{\twoonly}{\textbf{P2}\xspace}
\newcommand{\threeonly}{\textbf{P3}\xspace}
\newcommand{\fouronly}{\textbf{P4}\xspace}
\newcommand{\fiveonly}{\textbf{P5}\xspace}
\newcommand{\sixonly}{\textbf{P6}\xspace}
\newcommand{\sevenonly}{\textbf{P7}\xspace}
\newcommand{\eightonly}{\textbf{P8}\xspace}
\newcommand{\nineonly}{\textbf{P9}\xspace}
\newcommand{\tenonly}{\textbf{P10}\xspace}

\newcommand{\twelveonly}{\textbf{P12}\xspace}
\newcommand{\thirteenonly}{\textbf{P13}\xspace}
\newcommand{\fourteenonly}{\textbf{P14}\xspace}
\newcommand{\fifteenonly}{\textbf{P15}\xspace}
\newcommand{\sixteenonly}{\textbf{P16}\xspace}
\newcommand{\seventeenonly}{\textbf{P17}\xspace}

\newcommand{\four}{\textbf{P4} (Economist)\xspace}

\newcommand{\sixteen}{\textbf{P16} (ML Engineer)\xspace}

\newcommand{\onepp}{\textbf{P1} (HCI Post-Doc, PP=2)\xspace}
\newcommand{\twopp}{\textbf{P2} (Economist, PP=6)\xspace}
\newcommand{\threepp}{\textbf{P3} (Economist, PP=10)\xspace}
\newcommand{\fourpp}{\textbf{P4} (Economist, PP=8)\xspace}
\newcommand{\fivepp}{\textbf{P5} (PETs Practitioner, PP=8)\xspace}
\newcommand{\sixpp}{\textbf{P6} (Data Scientist, PP=10)\xspace}
\newcommand{\sevenpp}{\textbf{P7} (Prof. Data Policy, PP=4)\xspace}
\newcommand{\eightpp}{\textbf{P8} (Cryptographer, PP=6)\xspace}
\newcommand{\ninepp}{\textbf{P9} (HCI Researcher, PP=8)\xspace}
\newcommand{\tenpp}{\textbf{P10} (PETs Ed. Fellow, PP=6)\xspace}
\newcommand{\elevenpp}{\textbf{P11} (Viz. Researcher, PP=8)\xspace}
\newcommand{\twelvepp}{\textbf{P12} (Psych. Researcher, PP=1)\xspace}
\newcommand{\thirteenpp}{\textbf{P13} (Pub. Policy Researcher, PP=6)\xspace}
\newcommand{\fourteenpp}{\textbf{P14} (ML/Policy Researcher, PP=10)\xspace}
\newcommand{\fifteenpp}{\textbf{P15} (Prof. Med., PP=2)\xspace}
\newcommand{\sixteenpp}{\textbf{P16} (ML Engineer, PP=2)\xspace}
\newcommand{\seventeenpp}{\textbf{P17} (Technical PM, PP=2)\xspace}

\newcommand{\dataexperts}{data experts\xspace}
\newcommand{\Dataexperts}{Data experts\xspace}
\newcommand{\DataExperts}{Data Experts\xspace}

\newcommand{\RQone}{\textbf{RQ1}\xspace}
\newcommand{\RQtwo}{\textbf{RQ2}\xspace}

\newcommand{\lawq}{``Number of participants expressing explicit existing data privacy related legal concerns, or the need for updated laws/regulations.''}

\newcommand{\needq}{``Number of participants expressing needs or desires for privatized data (as opposed to encrypted/otherwise secured data).''}

\newcommand{\skepticismq}{``Number of participants expressing skepticism of existing differentially private data privacy methods.''}

\newcommand{\censusq}{``Number of participants that mention the U.S. Census.''}

\newcommand{\validationq}{``Number of participants that mention a need or desire for validating differentially private outputs against real data.''}

\newcommand{\syntheticq}{``Number of participants with concrete experience working with synthetic data.''}

\newtheorem{theorem}{Theorem}
\newtheorem{definition}[theorem]{Definition}

\usepackage{pdfrender}
\newcommand*{\boldcheckmark}{%
  \textpdfrender{
    TextRenderingMode=FillStroke,
    LineWidth=1.0pt, 
  }{\checkmark}%
}

\begin{document}

\title[Are Data Experts Buying into Differentially Private Synthetic Data?]{Are Data Experts Buying into Differentially Private Synthetic Data?\\ Gathering Community Perspectives}


\author{Lucas Rosenblatt}
\affiliation{%
  \institution{Tandon School of Engineering, New York University}
  \city{New York}
  \state{NY}
  \country{USA}
  }
  \author{Bill Howe}
\affiliation{%
  \institution{Information School, University of Washington}
  \city{Seattle}
  \state{WA}
  \country{USA}
  }
  \author{Julia Stoyanovich}
\affiliation{%
  \institution{Tandon School of Engineering and Center for Data Science, New York University}
  \city{New York}
  \state{New York}
  \country{USA}
  }

\renewcommand{\shortauthors}{Rosenblatt et al.}

\begin{abstract}
Data privacy is a core tenet of responsible computing, and in the United States, differential privacy (DP) is the dominant technical operationalization of privacy-preserving data analysis. With this study, we qualitatively examine one class of DP mechanisms: private data synthesizers. To that end, we conducted semi-structured interviews with \textit{data experts}: academics and practitioners who regularly work with data. Broadly, our findings suggest that quantitative DP benchmarks must be grounded in practitioner needs, while communication challenges persist. Participants expressed a need for context-aware DP solutions, focusing on parity between research outcomes on real and synthetic data. Our analysis led to three recommendations: (1) improve existing insufficient sanitized benchmarks; successful DP implementations require well-documented, partner-vetted use cases, (2) organizations using DP synthetic data should publish discipline-specific standards of evidence, and (3) tiered data access models could allow researchers to gradually access sensitive data based on demonstrated competence with high-privacy, low-fidelity synthetic data. 
\end{abstract}




\received{20 February 2007}
\received[revised]{12 March 2009}
\received[accepted]{5 June 2009}

\maketitle

\section{Introduction}
\label{sec:intro}

Differential privacy (DP) provides a principled and broadly applicable approach for enabling access to data that would otherwise be too sensitive to share~\cite{dwork2014algorithmic, machanavajjhala2017differential,liu2021machine}. While direct applications of DP add noise to a specific statistical result, DP data synthesis mechanisms model the original data distribution, inject noise, then sample the noisy model to generate a synthetic dataset that satisfies differential privacy: the dataset is about as likely to have been generated with or without any particular row in the original data.  These synthetic datasets are then intended to be used as a drop-in replacement for the original, with the expectation that any statistical computation over the synthetic data will produce a sufficiently similar result. These \textit{differentially private synthetic data} approaches are increasingly studied in both methodology \cite{zhang2017privbayes, hardt2010simple, liu2021iterative, aydore2021differentially} and evaluation \cite{arnold2020really, tao2021benchmarking, rosenblatt2023epistemic}, and are included explicitly in the definition of privacy-enhancing technologies in the 2023 White House Executive Order on the Safe, Secure, and Trustworthy Development and Use of Artificial Intelligence~ \cite{biden2023executive}.

Evidence to support claims of utility is typically presented through proxy tasks over benchmark datasets (\eg the ubiquitous Adult dataset~\cite{adult}).  Proxy tasks may include descriptive statistics, queries involving one or two variables~\cite{hill2015evaluating,hay2016principled,takagi2021p3gm,tao2021benchmarking}, classification accuracy~\cite{ding2020differentially,takagi2021p3gm,zhang2017privbayes}, and information theoretic measures~\cite{zhang2017privbayes}. Although these proxy tasks are procedurally representative of real tasks, the implicit claim of generalization to practice is rarely examined.

Limited empirical evidence on relevant tasks undermines trust in the practical use of DP.  The US Census Bureau adopted DP for disclosure avoidance in the 2020 census, interpreting federal law (the Census Act, 13 U.S.C. § 214, and the Confidential Information Protection and Statistical Efficiency Act of 2002) as a mandate to use advanced methods to protect against computational reconstruction attacks unforeseen when the laws were passed.  
But the adoption of DP for the Census was met with resistance among many in the research community, who contend that data infused with DP noise affects demographic totals~\cite{boyd_2022,ruggles2019differential} and exacerbates underrepresentation of minorities~\cite{kenny2021use, ganev2021robin}.  Besides the research implications, there are potential consequences for policy: In the US, block grants are allocated based on minority populations as measured by the census data, and underrepresentation can lead to underfunding integral services including Medicaid, Head Start, SNAP, Section 8 Housing vouchers, Pell Grants, and more~\cite{christ2022differential}.  Although the Census Bureau held workshops, released demonstration datasets, and published technical reports to support the community, these outreach efforts realized limited success; multiple lawsuits are still pending as of September 2024.
This context suggests that we do not understand how researchers, technologists, and practitioners—the target audience for differentially private synthetic data—engage with epistemological questions of utility, risk, evidence, and trust in the use of this DP synthetic data.
\begin{definition}[Data Expert] \label{def:data_expert} We refer to \dataexperts, who are individuals that, in a professional capacity, engage directly with data through activities such as collection, management, analysis, or interpretation. Data experts typically possess specialized knowledge or skills in handling data and may occupy the following roles: \textbf{researchers} who conduct empirical studies requiring data-driven methodologies, \textbf{data scientists} who process and analyze large datasets to extract insights, \textbf{practitioners} in fields like economics, medicine, education, public policy, and psychology, where data informs practice and decision-making, or \textbf{policy makers} who rely on data to shape policies and recommendations.
\end{definition}

\subsection{Overview of methodology} 

We conducted an IRB-approved study to interview 17 \dataexperts about the risks and benefits of using DP synthetic data. Our study participants represented a variety of experience levels, including Ph.D. students, senior faculty members, and professionals, and were from fields that rely heavily on sensitive data, including economics, medicine, education, data science, public policy, and psychology.  Through a semi-structured interview, we elicited participants' perspectives on privacy in general (to contextualize their responses with respect to their ``privacy prior'') and about specific approaches to protecting privacy for data release.
We also asked them to consider a hypothetical scenario involving differentially private medical data. 
We designed the study to answer two main research questions:
\begin{enumerate}
    \item[\RQone] \textbf{(Current and Future Utility)} In what contexts do \dataexperts create, share, and use differentially private synthetic data in their work? What benefits do \dataexperts perceive in its potential use? 
    \item[\RQtwo] \textbf{(Concerns, Limitations, and Suggestions)} What concerns do \dataexperts have in using differentially private synthetic data in place of original data?  What limitations do they experience in practice?  What suggestions do they have to make differentially private synthetic data useful to them?
\end{enumerate}

\subsection{Summary of contributions} 
As a by-product of answering \RQone and \RQtwo, we gathered insights into our participants' \textit{general} understanding of data privacy. Participants had varying levels of familiarity with the technical foundations of data privacy, but all demonstrated appreciation and understanding of the societal and philosophical context motivating the technology, and most underscored the importance of considering this context when implementing DP and evaluating the trade-offs.  
(\RQone) Most participants reported that they do not \emph{currently} use differentially private synthetic data, and would only consider using it as a last resort.  
(\RQtwo) Participants expressed ambivalence in weighing the value of privacy guarantees against the epistemological ramifications, emphasizing the risk of incorrect decisions being made about specific individuals or underrepresented groups as a result of DP noise.
All participants expressed that validation against real data is required, though with little consensus about the form this validation should take. All participants saw the opportunity of broadening access to sensitive data to which researchers otherwise would not have access.  Most participants recognized the need for authoritative guidance and regulatory oversight, since incorrect implementations (or misuse of correct implementations) could lead to the worst of both worlds: realistic data that masquerades as both privacy-preserving and source-faithful, yet suppresses important correlations or introduces spurious ones, leading to false conclusions. 

\subsubsection*{Recommendations} Our analysis and discussion led us to the following three recommendations:
\begin{enumerate}
    \item[]\textbf{\textit{(Recommendation 1)}}~~DP researchers should commit to providing \textbf{evidence of validation} in at least \textbf{one partner-vetted use case}. Sanitized public benchmarks, divorced from their original use case and cleansed to emphasize expected correlations at the expense of unexpected challenges (missing data, rare classes, spurious correlations), are insufficient.
    \item[]\textbf{\textit{(Recommendation 2)}}~~Organizations consuming differentially private data should agree on, and publish, their \textbf{standards of evidence}.  In many cases, these norms will be aligned with the shared training of the organization's members: \textbf{statisticians may demand precise characterization} of the error introduced by the privacy mechanism, while \textbf{empiricists may develop application-specific benchmarks}.
    \item[]\textbf{\textit{(Recommendation 3)}}~~We recommend \textbf{tiered access to sensitive data}, where promising results on (and demonstrated safe handling of) high-privacy but low-fidelity data can be used to apply for greater access to higher-risk differentially private data, or to the original data itself.  This \textbf{``driver's license''} model acknowledges the ad hoc, exploratory, and iterative nature of research, especially research in small labs where stringent, security-oriented privacy protocols are difficult to apply reliably.
\end{enumerate}

\subsection{Paper organization} 
We start with an overview of related work in Section~\ref{sec:related}, and then go on to describe study design in Section~\ref{sec:methods}.  We present findings and results from our interviews in Section~\ref{sec:results}, and contextualize, discuss and analyze these findings in Section~\ref{sec:discussion}. We conclude with our recommendations in Section~\ref{sec:conc}.
\section{Related Work}
\label{sec:related}
In this section we first provide the definition of differential privacy (for completeness) and discuss the field of DP synthetic data at large. Then, we highlight related work from a \textit{data subject} perspective; this is followed by a section on more closely related work that details perspectives of \textit{\dataexperts}. Note that the \textit{most} closely related work to our study, and how it differs from our results, is summarized in Table~\ref{tab:rel_work}.

\subsubsection*{Background on Differential Privacy} Differential privacy (DP) is a formal privacy guarantee limiting the amount of information a statistical release mechanism reveals about any one individual in the dataset. In ensuring minimal privacy loss, DP mechanisms introduce carefully calibrated \textit{noise} into statistics and mechanism outputs. We provide the formal definition of DP here, for completeness. Let $\mathcal{D}$ denote a data universe and $\mathcal{D}^n$ represent the space of datasets of size $n$. Two datasets $D, D' \in \mathcal{D}^n$ are considered neighboring, denoted $D \sim D'$, if they differ by only a single record.

\begin{definition}[Differential Privacy \cite{dwork2014algorithmic}]\label{def.dp}
Mechanism $\mathcal{M}: \mathcal{D} \rightarrow  \mathbb{R}$ is \emph{$(\epsilon,\delta)$-differentially private} if for every pair of neighboring databases $D,D' \in \mathcal{D}$, and all possible subsets of outputs $\mathcal{S} \subseteq \mathbb{R}$,
\[\Pr[\mathcal{M}(D) \in \mathcal{S}] \leq \exp(\epsilon)\Pr[\mathcal{M}(D') \in \mathcal{S}]+\delta. \]
\end{definition} 
The DP framework is robust against current and future attacks and by design auxiliary information does not provide adversaries with any large advantages. The precise amount of noise provides a quantifiable guarantee for privacy loss, allowing data collectors and analysts to make informed decisions about the trade-offs between \textit{data utility} and \textit{privacy protection}. Tuning the privacy parameters $\epsilon$ and $\delta$ balances such trade-offs, with smaller values indicating stronger privacy guarantees. For these reasons and more, DP has emerged as a gold standard data privacy definition, and has been used widely in the fields of data science and machine learning~\cite{dwork2008differential,zhao2022survey, blanco2022critical,ji2014differential}. Despite its advantages, we will discuss the complexity involved in \textit{practically} implementing DP; successful deployments require a careful consideration of data context, the selection of the appropriate mechanism, and noise calibration. To that end, many challenges remain in its successful integration into the social sciences~\cite{oberski2020differential} and medical research~\cite{dankar2013practicing, ficek2021differential}. 

\begin{table*}[b]\label{tab:rel_work}
\centering
\caption{Summarizing related survey and interview studies on deployment challenges and opportunities for Differential Privacy (DP) and Synthetic Data (SD). To the best of our knowledge, we offer the first \textbf{interview study} assessing perspectives on the use of \textbf{differential privacy} for \textbf{synthetic data}.}
\resizebox{\linewidth}{!}{
\begin{tabular}{m{3cm}m{3cm}m{1.5cm}m{1.5cm}m{1.5cm}m{7cm}}
\midrule
\textbf{Reference} & \textbf{Study Details} & \textbf{Interviews?} & \textbf{DP Focus?} & \textbf{SD Focus?} & \textbf{Summary} \\
\midrule
\midrule
boyd et. al \cite{boyd_2022} & \textbf{Interviews}, n=47, census stakeholders & $\checkmark$ & $\checkmark$ &  & (i) Epistemic disagreements (ii) Politically and economically charged setting \\
\midrule
Williams et. al \cite{williams_etal} & \textbf{Survey}, n=1,000+ AEA members &  & $\checkmark$ & $\checkmark$ & (i) Division on privacy vs. utility (ii) Lack of familiarity with DP \\
\midrule
Dwork et. al \cite{dwork2019differential} & \textbf{Interviews}, n=11, DP practitioners from 7 organizations & $\checkmark$ & $\checkmark$ &  & (i) Lack of consensus on $\epsilon$ (ii) Implementation decision challenges \\
\midrule
Garrido et. al \cite{DBLP:journals/popets/GarridoLMS23} & \textbf{Interviews}, n=24, industry practitioners & $\checkmark$ & $\checkmark$ &  & (i) Practical gaps in tooling (ii) Misalignment with business priorities \\
\midrule
Bullek et. al \cite{DBLP:conf/chi/BullekGMP17} & \textbf{Survey}, n=228, MTurkers, general population &  & $\checkmark$ &  & (i)  Measured trust in Randomized Response (ii) Advocate simple privacy explanations \\
\midrule
Sarathy et. al \cite{sarathy2023don} & \textbf{Interviews}, n=19, DP Creators (OpenDP) & $\checkmark$ & $\checkmark$ &  & (i) Practitioners had limited understanding of DP (ii) Clashes at every stage of data science workflow \\
\midrule
van Hoorn et. al \cite{van2024acceptance} & \textbf{Interviews}, n=24, data scientists, medical professionals & $\checkmark$ &  & $\checkmark$ & (i) Assessed uptake of SD in medical profession (ii) Utility of SD in ML-based healthcare innovation \\
\midrule
Ngong et. al \cite{ngong2023evaluating} & \textbf{Interviews}, n=24, data practitioners & $\checkmark$ & $\checkmark$ &  & (i) Recommendations to improve existing DP tools’ usability (ii) Recount participants' experiences \\
\midrule
Steed et. al \cite{steed2024adoption} & \textbf{Interviews}, n=28, privacy preserving analysts (PPA) & $\checkmark$ & $\checkmark$ &  & (i) Develop grounded theory of PPA adoption (ii) Recount design justifications \\
\midrule
\textbf{This work} & \textbf{Interviews,~n=17,~\dataexperts} & $\boldcheckmark$ & \textbf{$\boldcheckmark$} & \textbf{$\boldcheckmark$} & \textbf{(i) \Dataexperts agree on importance of data privacy, but have concerns about generalizability of results - this prevents uptake of DP synthetic data in practice - respondents considered it as last resort despite clearly articulating the potential benefits (ii) Leads to calls for partner-vetted use cases, public standards of evidence, and a ``driver's license'' model for tiered data access} \\
\bottomrule
\end{tabular}
}
\label{tab:rel_work}
\end{table*}

\subsubsection*{Perspectives of the Data Subjects} Attitudes toward data privacy among data subjects have been investigated extensively in recent years~\cite{DBLP:conf/chi/BullekGMP17,DBLP:conf/ccs/CummingsKR21,kacsmar2023comprehension,DBLP:conf/soups/KuhtreiberPR22,nanayakkara2023chances,sannon2018privacy,xiong2020towards,xiong2022using,xiong2023exploring} (i.e. the individuals who comprise the data and their ability to understand DP guarantees and parameters).
Most (although not all) authors find that data subjects' willingness to share data in a DP regime is mediated by explanations of DP methods and associated parameters. Bullek et al., present a survey of 228 participants about trust in randomized response (a simple DP technique), finding that demonstrating the effect of noise  increased trust, but that decision-making was only influenced after providing ``simple explanations of usable privacy'' ~\cite{DBLP:conf/chi/BullekGMP17}.  
Nanayakkara et al. found that odds-based explanations (as opposed to example-based or intuitive explanations) mediate respondents' willingness to share data based on changing $\epsilon$ values based on a survey of 963 individuals~\cite{nanayakkara2023chances}.  However, Smart~\etal, using large-scale online surveys, found that explanations have little effect on willingness to share~\cite{smart2022understanding}.

These insights have motivated the development of visualization methods to communicate uncertainty and risk~\cite{DBLP:conf/chi/BullekGMP17,nanayakkara2023chances}, and spurred work on explaining technical methods through concrete scenarios~\cite{kacsmar2023comprehension,xiong2020towards}, high-level metaphors~\cite{karegar2022@exploring} and illustrations~\cite{xiong2022using,xiong2023exploring}.
In contrast to this work, we target \emph{\dataexperts}: those with responsibility and training to produce or consume sensitive data in order to inform decisions related to public or private interests; this is a primary target audience for differential privacy. In Table~\ref{tab:rel_work}, we compare and contrast closely related studies to ours that engage with similar research questions among similar groups of participants; we do this to highlight the main novelty of our work, \textbf{an interview study that analyzes perspectives on differentially private \textit{synthetic data} as a DP use case}. In the following section, we expand on findings from previous work. We contextualize our findings with respect to these prior studies when discussing results in Section~\ref{sec:discussion}.

\subsubsection*{Perspectives of the \DataExperts}
The U.S. Census Bureau implemented DP in the 2020 Decenial census, leading to significant academic and legal debate.  
Authors boyd and Sarathy~\cite{boyd_2022} examine the controversy through the lens of epistemic disconnects between different perspectives informed by an interview study of 47 census stakeholders: ``Debates about data quality appear to be about technical matters, but they also involve unacknowledged epistemic disagreements that undermine efforts to reconcile technical claims.''  We corroborate the finding that different epistemic perspectives mediate opinions about privacy technology, but our interview procedure was designed to surface these perspectives objectively, de-emphasizing the politically charged setting of census policy.

Williams~\etal conducted a survey of over 1,000 members of the American Economic Association (AEA) to study their familiarity with DP and their attitudes towards its use \cite{williams_etal}. The authors found that economists, even those working with Federal micro data, are typically unfamiliar with DP and that they are divided on whether privacy protections justify degradation in utility.  The authors also collected information about the types of data analysis that are frequently conducted from the respondents, and about how they would control the privacy / utility trade-off across these tasks.

Dwork~\etal interviewed 11 DP practitioners from 7 organizations with substantial technical expertise in privacy, and found a broad lack of consensus among respondents in choosing $\epsilon$ and other key DP implementation decisions \cite{dwork2019differential}.  

Garrido~\etal  interviewed 24 industry practitioners, with different levels of familiarity with DP, to understand the business case for privacy in their organizations, and the challenges they are facing in deploying DP \cite{DBLP:journals/popets/GarridoLMS23}.  They found that, while some organizations are using DP, most are not, due to the practical gaps in tooling (\eg the author found that ``Running analysts’ scripts without `seeing' the data is a distant reality for the interviewed companies.''), and to the frequent misalignment between the promise of DP and the companies' business priorities.

Several usability studies of DP tools have also been conducted in recent years~\cite{murtagh2018usable,ngong2023evaluating,sarathy2023don,zhu2023making}.  Sarathy~\etal  interviewed 19 data owners who are using a DP data analysis prototype to release privacy-preserving data statistics \cite{sarathy2023don}.  They found that while DP can expand public access to data, facilitate exploratory research, and improve replication, it also disrupts existing workflows and introduces evidentiary risk due to lack of experience among users. Ngong~\etal  interviewed 24 U.S. data practitioners regarding their use of four open-source DP tools ~\cite{ngong2023evaluating}.  They found that DP tools can help novice users learn about DP, and recommend to improve the usability of the tools for broader adoption. Additionally, Hod~\etal conducted a large scale study of privacy in practice, releasing synthetic microdata of live births in
Israel, and attempted to follow best DP practice throughout \cite{hod2024differentially}.

\subsubsection*{General Data Privacy Studies}\label{sec:gen_priv_study} A rich body of interview studies has assessed general attitudes towards data privacy across various communities and groups; however, these studies do not focus on differential privacy (DP) or synthetic data (SD). Some interview studies have focused on software practitioners and/or app developers, revealing data privacy perceptions, attitudes and behaviours specific to these roles \cite{balebako2014privacy, hadar2018privacy, li2018coconut, gutfleisch2022does, iwaya2023privacy, kekulluouglu2023we}. Other studies have considered motivations of executives and business leaders in enacting privacy interventions for legal and regulatory compliance, with frameworks such as the GDPR considered \cite{sirur2018we, bednar2019engineering, emami2020ask, dalela2022study}. A particularly relevant study interviewed  ``privacy champions'' i.e. professional privacy evangelists in various roles, discussing their implementation and advocacy strategies \cite{tahaei2021privacy}. As a whole, these studies help understand general attitudes towards ``data privacy'' without specifically investigating differential privacy and synthetic data as primary implementation strategies.

\subsubsection*{\Dataexperts} Unlike data subjects, for whom privacy protection is the primary concern, \dataexperts are invested primarily in the reliability and generalizability of the conclusions drawn from the data, and are therefore naturally hesitant to adopt procedures that may undermine those conclusions. 
In this work, we focus specifically on private data synthesis as an interface between diverse collection contexts and diverse application contexts, asking respondents to reflect on specific scenarios and the evidentiary standards they would accept in those scenarios.

\section{Research Methodology}
\label{sec:methods}

\begin{table*}
    \centering
    \caption{Details on each participant. Participants were asked about the domain of their work and their education level. Additionally, we asked participants to define three data privacy terms to assess their familiarity with data privacy concepts, constructing a numeric index called \textit{Privacy Prior} (PP). Terms were ``de-identified data'' (+2 for complete definition, +1 for partial), ``k-anonymity'' (+4 for complete definition, +2 for partial), and ``differential privacy'' (+4 for complete definition, +2 for partial); the Privacy Prior score serves as a general indicator of familiarity with data anonymization and obfuscation tools. }
    \resizebox{\textwidth}{!}{
    \begin{tabular}{llllr|llllr}
    \toprule
    ID & Sector & Domain/Expertise & Education & PP & ID & Sector & Domain/Expertise & Education & PP \\
    \midrule
    P1 & Academic & HCI Researcher & PhD & 2 & P10 & Acad./Govt. & Privacy Educator & MA & 6 \\
    P2 & Academic & Economist & PhD & 6 & P11 & Academic & Visualization (HCI) & PhD & 8 \\
    P3 & Acad./Govt. & Census (Retired) & PhD & 10 & P12 & Academic & Clinical Psych. & PhD & 1 \\
    P4 & Academic & Economist & PhD & 8 & P13 & Academic & Pub. Policy & PhD & 6 \\
    P5 & Acad./Govt. & Edu. Researcher & PhD & 8 & P14 & Academic & Pub. Policy/ML & PhD & 10 \\
    P6 & Industry & Data Scientist & MA & 10 & P15 & Academic & Prof. of Medicine & PhD & 2 \\
    P7 & Acad./Govt. & Library Sciences & PhD & 4 & P16 & Industry & NLP Engineer & MA & 2 \\
    P8 & Industry & Cryptographer & PhD & 6 & P17 & Industry & Technical PM & BA & 2 \\
    P9 & Academic & HCI Researcher & PhD & 8 &  &  &  &  &  \\
    \bottomrule
    \end{tabular}
    }
    \label{tab:participants}
\end{table*}

\subsubsection*{Participants and Recruitment}
\label{sec:recruitment}
Participants were initially recruited via public postings (through list serves and Slack channels) related to data privacy and synthetic data. In order to interview \dataexperts from outside these technical fields, we also recruited through our professional network (using a snowball sampling strategy). Details on each participant can be found in Table~\ref{tab:participants}. All participants were based in the U.S. at the time of the study. Additional aggregate race and demographic information can be found in Table~\ref{tab:summary_fractions} in Section~\ref{app:dem} of the appendix. We recruited medical professionals, social scientists, and technical industry professionals. The majority of our participants did not work directly on data privacy, but all at minimum worked closely with data in their respective fields. Participants represented a variety of experience levels, from PhD students to senior faculty members and retired US Census professionals. 

\subsubsection*{Saturation and Sample Size}
We grew our study with the goal of achieving \textit{thematic saturation} (the point at which no new themes or codes emerge from additional interviews) with regards to \RQone and \RQtwo \cite{guest2006many, fusch2015we}. This is a standard practice in qualitative interview studies; we found that after conducting 17 interviews with participants from diverse professional backgrounds, we observed recurring themes and consistent patterns in codes related to contexts of DP synthetic data use and practical concerns, limitations and suggestions. Of course, diversity in participants' expertise and backgrounds naturally introduces variability when deciding saturation had been reached \cite{guest2020simple}. While additional interviews may provide more nuanced insights, we concluded, given the depth and richness of the data we had already collected, that the likelihood of new participants altering the established themes related to \RQone and \RQtwo was minimal, hence reporting on our 17 interviews.%
\newline

\subsubsection*{Semi-Structured Interview Methodology}
\label{sec:methodology}
We employed semi-structured interviews to understand perspectives on privacy and non-privacy-related technical topics. The average duration of the interviews was 47 minutes, with the longest being 1 hour and 6 minutes and the shortest 41 minutes. Interviews were structured around a set of predefined questions (See Appendix~\ref{app:questions} for complete materials), although participants were encouraged to share any details from their experiences that they deemed relevant. 

\begin{figure*}
\includegraphics[width=\textwidth]{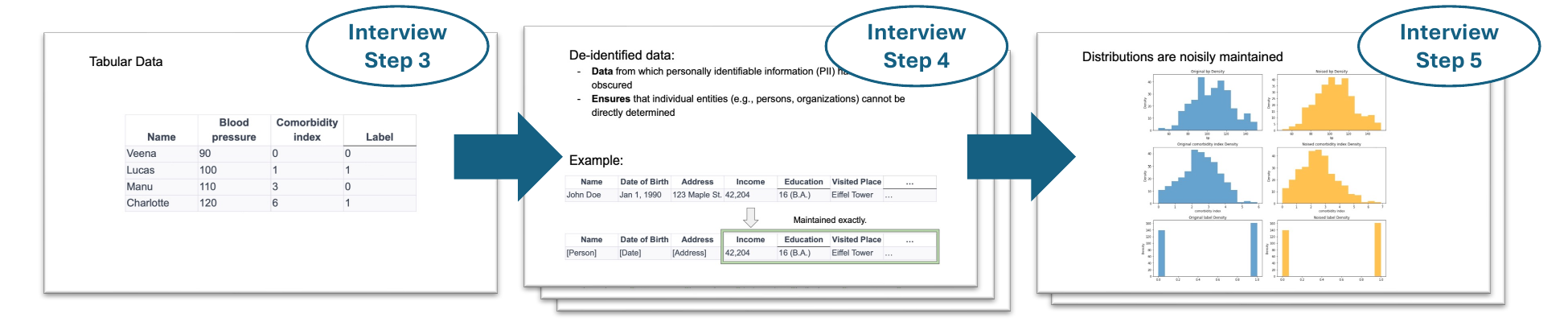}
\caption{Materials used in Steps 3--5 of the interview. In the first part of the slide prompts, participants were presented with sample data and asked about it's sensitivity. In the second part, they were presented with privacy terms to define, and then were given a definition. In the third part, they were shown histogram comparisons and scatterplot correlations between the real and fake data.}
\label{fig:slides}
\end{figure*}

Our interviews were structured as follows. \textit{Step 1.} An initial set of ``Who are you?'' questions was asked, designed to gather basic information (self-described identify and domain of work). \textit{Step 2.} We transitioned into privacy related questions using ``What does the term `data-privacy' mean to you?'' as a jumping off point, followed by probing questions to better understand their response. \textit{Step 3.} We switched into visual prompts and asked about thoughts on sensitive data (using a slide with an example of tabular medical data). \textit{Step 4.} To assess their prior familiarity with data privacy, we asked for ``in-your-own-words'' definitions of three data privacy terms (de-identified data, k-anonymity and differential privacy), using visual prompts on slides. \textit{Step 5.} After the privacy term definitions, we entered into a hypothetical scenario involving the same tabular medical sample shown earlier, alongside a differentially privatize, synthetic version of that data. We showed participants basic histograms and correlations on this data and asked for their thoughts, allowing room for an open-ended discussions. 
Figure~\ref{fig:slides} shows the slides we used to support Steps 3-5 of the interview. \textit{Step 6.} We followed up with any ``blue-sky'' questions (targeting the participants background and post-interview reactions to differentially private synthetic data), or prompted the participant to expand on previous statements from earlier sections considered interesting or relevant by the interviewer.

We note here that, to assess participants' prior familiarity with data privacy, we asked for ``in-your-own-words'' definitions of three data privacy terms (de-identified data, k-anonymity and differential privacy), using visual prompts on slides. This helped us ascertain the extent of a participants' general familiarity with  data privacy concepts (they were allowed to simply say they were unfamiliar with these terms). The exact calculation of the ``Privacy Prior'' (PP) score is given in the caption for Table~\ref{tab:participants}; this score provided us with a rough indicator of participant familiarity with the data privacy space at large. 

\subsubsection*{Coding and Analysis}
We conducted a thematic analysis (as outlined by Braun et al., Maguire et al. \cite{braun2006using, maguire2017doing}). One researcher began with open coding a small set of interviews using the Atlas.ti \cite{ATLASti2023} software suite. Then, multiple researchers reviewed the codes to reconcile inconsistencies, and together applied the resulting codes to the entire set of interview transcripts.
Through this process, a two-tier heirarchical thematic coding was identified: in the top tier, a set of 7 high-level themes were identified. The second tier then contained 35 more granular codes, which could be attached to responses. See Section~\ref{app:codebook} in the Appendix for the complete list of themes and codes, each accompanied by an example associated response.
We do not report inter-rater reliability (IRR), following the guidance of \citeauthor{mcdonald2019reliability} \cite{mcdonald2019reliability}, as coding was integral to our process of qualitative analysis rather than a by-product of the study. Moreover, the researcher who developed the initial themes and codes is an expert in data privacy; as they also conducted the interviews, they took on the role of ethnographer throughout the data collection and initial coding process.
Our coding allowed us to extract semantic and latent themes from the interviews that directly related to our research questions, as well as interviewee concurrence and representative quotations for direct inclusion in our discussion. Additionally, the granularity of the coding allowed us to generate Figure~\ref{fig:percent} and the summary tables in Section~\ref{sec:results}.

\section{Findings}
\label{sec:results}
\subsection{Participants Backgrounds on Data Privacy - General Findings} 
In our interviews, participants expressed general beliefs about data privacy and differential privacy (both prompted and unprompted) that provide the necessary context for our insights into \RQone and \RQtwo. 

\begin{table*}[h!]
 \centering
 \caption{Summarizing \Dataexperts' understandings of technical \& philosophical perspectives on data privacy.}
  \begin{tabular}{|p{2.5cm}|p{12cm}|}
    \hline
    \rowcolor[gray]{0.9} \textbf{Theme} & \textbf{Explanation} \\
    \hline
    Holistic Data\newline Privacy & Participants emphasized that data privacy, properly implemented, is comprehensive, considering the entire data lifecycle and protecting data both at the individual and aggregate levels. \tenpp: \textit{``Data privacy means ensuring that the right systems and design is in place  ...  how it's stored and how it's moved ... It means adequate protection of the individuals or institutions ... 
    both at a microdata level as well as at the aggregate level.''}  \\
    \hline
    \rowcolor[gray]{0.9} Domain Specific Data Privacy & Nearly all respondents noted that the requirements of data privacy vary across domains, nuanced by context of collection and use. \sevenpp: \textit{``In healthcare, I expect one level of privacy, and with social media, I expect a different level of privacy.''} \\
    \hline
    Data Privacy as\newline Cultural Artifact & Data privacy technology can center and reify predominantly Western ideas of individualism
    \seventeenpp: \textit{``[In] some ... countries around the world, privacy concerns [are] pushed aside in certain areas for the public good.''} \\
    \hline
    \rowcolor[gray]{0.9} Teaching, Communication and Instilling Best Practice & Effectively communicating data privacy concerns and best practices (with both experts and non-experts) fosters adoption. \eightpp acknowledges that aspects of data privacy and differential privacy are \textit{``very hard to interpret,''} and \onepp and \threepp stress the importance of establishing protocols and practicing transparency \textit{``in good faith''} with all stakeholders. \\
    \hline
  \end{tabular}
 \label{tab:rq1}
\end{table*}

Before delving into the participant responses, we note that, on average, study participants had at least some privacy expertise (average Privacy Prior $= 5.8 / 10$, median $= 6 / 10$, see Table~\ref{tab:participants} for calculation methodology). Their conceptions of ``data privacy'' varied, even among \dataexperts with high privacy priors ($\geq 6$). For example, when asked: ``How would you characterize the term \textit{data privacy}?'' \ninepp stated that ``Everyone has a  different definition of privacy ... it depends on the data owner,'' going on to say that ``Privacy means that if data is collected for one purpose ... then it should not be used for any other purpose.''

\subsubsection*{Holistic Data Privacy} \eightpp, \tenpp\footnote{PETs stands for \textbf{P}rivacy \textbf{E}nhancing \textbf{T}echnologies}, and \sixteen shared holistic perspectives when defining ``data privacy.'' As \tenonly put it, ``If statistics are released on a data set, those individuals [should be] protected ... from the beginning [of collection] until the release at the public end.'' \eightonly summarized the practice of privatizing outputs or data as ``the control of [any] information that flows through [and the prevention] of information flow or leakage that's not allowed.'' \sixteenonly listed out important questions to keep in mind when considering the data privacy of a system: ``What is the data that you are generating by ... going [about] your life? How much of that is actually recorded? And then, ... do we have any autonomy over where that data goes? ... What are the end users of that data? [Are we] aware of our rights for that? Data privacy encapsulates all of that.''

\subsubsection*{Domain-specific Data Privacy} Many participants agreed that privacy definitions must be tailored to the  application domain. \sevenpp: ``In healthcare, I expect one level of privacy, and with social media, I expect a different level of privacy.'' \fourpp considered an even finer-grained situational nuance: `` ... seeing names as part of data that is published doesn't by itself make it an infringement of privacy ... it really depends on the socio-economic, legal context of how the data was collected, how it is being disseminated, and what the intended use of it is,'' and goes on to summarize with ``What is one person's private data might be another person's public persona.''

\subsubsection*{Cultural Differences} Prior work has noted the narrow view of Western individualism in data privacy formalisms \cite{bailey2007seizing, basu2012privacy}. Aligning with this perspective, \seventeenpp observes that ``[In] some ... countries, privacy concerns [are] pushed aside ... for the public good,'' citing Sweden's census initiative collecting private data and allowing researcher access. ``That would never be allowed in the United States.'' He adds, ``It's the same case in Singapore [and other] eastern cultures.'' However, most of our U.S.-based participants did not mention cultural priors or question an individual's right to privacy. Intra-cultural differences, like generational attitudes toward data comfort, are also often neglected. \sevenpp: ``I know people ten or more years older than me who say, 'Oh, the kids don't care about privacy, they put everything on social media.' ... [But] students absolutely care ... So we need to continue ... to respect other people [and their attitudes] as well.''

\subsubsection*{Challenges in Data Privacy Protocols}
Data privacy faces an uphill battle in many research or industry settings, where protocol is unclear or unestablished. For example, \twelvepp notes that in her lab, ``People are not clear on ... how to keep data private or how to share data or [even] where can data be stored?'' Breaches may occur due to ``a miscommunication ... or a lack of a clear protocol.'' She admits, ``there're still times where [I ask], 'Can I have access to this? How should I store this? How can I send this?'''

This uncertainty stems from a research priority attitude, which should be considered in data privacy protocol design. 
\fourpp summarizes a typically chaotic research process: ``There's a mental research process that researchers go through. `I'm gonna download a bunch of variables that I think are relevant, and then I'm gonna use a subset of those in my paper. I don't know how to write [the paper yet], and I don't know how to do the analysis, but I think [generally] this is what the papers about.''' To adequately support the use of sensitive data in need of privacy protection, these ad hoc, opportunistic processes should be recognized and supported (we attempt to accommodate them with our \textit{driver's license} model, see Sections~\ref{sec:intro}and~\ref{sec:discussion}).

Formalized protocols are more attainable in institutions with accountability regimes. 
\onepp discusses these challenges: ``... if I'm running an academic research project I will have gone through an Ethics Review ... there's definitely checks and balances. However, [as] a consultant ... there isn't a formal data agreement ... I just have free access to some of their documents and resources.''

\subsubsection*{Communicating Data Privacy} The need to better communicate data privacy emerged as a major theme. \fifteenpp noted that it's ``our responsibility to develop tools that protect privacy, [and] also ... to communicate ... [We want] to protect privacy and make people ... aware of the ... risks.'' \eightpp: ``[We] have to do a better job ... communicating [how we] protect privacy, that it's in [everyone's] interest ... to apply these techniques [of] differential privacy.''

Some participants, however, expressed skepticism that it is possible to explain data privacy to people lacking expertise in an actionable sense. \fourteenpp does not put ``much stock in ... lay person explanations,'' explaining that people might say, ``'Oh, yeah, that's cool. But [I have] a regular privacy concern, like I'm gonna get kicked out of my housing.''' Still, \eightpp found that with some technical background, using ``concrete metrics based on what attackers can actually do helps ... [in communicating] a concept like epsilon in differential privacy.'' Through ``very visual and very appealing'' presentations, they can ``[produce an explanation] that people understand ... they don't need that much background.'' Visual explanations of complex privacy parameters have been explored in recent work ~\cite{nanayakkara2023chances}, and are a promising avenue for future research. \ninepp expresses a need for tools to communicate privacy intuitions: ``It's so confusing, How do you select the proper epsilon value? What does [it] actually mean? ... What does [the epsilon value] mean for ... the data owner?''

\newcommand{\tightspacing}{\renewcommand{\baselinestretch}{0.8}\normalsize}

\begin{figure*}[t]
\begin{tikzpicture}
\begin{axis}[
    xbar,
    width=6cm,
    height=8cm,
    y axis line style = { opacity = 0 },
    axis x line       = none,
    tickwidth         = 0pt,
    legend style={at={(-2.1,0.3)}, anchor=north west},
    legend image code/.code={
            \draw[#1,fill=#1] (0cm,-0.1cm) rectangle (0.3cm,0.1cm);
        },
    yticklabel style={execute at begin node=\tightspacing, align=right,text width=10cm, font=\footnotesize},
    ytick             = data,
    enlarge y limits  = 0.2,
    symbolic y coords = {
        \censusq,
        \syntheticq,
        \validationq,
        \skepticismq,
        \needq,
        \lawq,
    },
    nodes near coords,
    nodes near coords align={horizontal},
    nodes near coords style={anchor=west},
    xtick={0,17},
    xticklabels={0,17},
    point meta=explicit symbolic,
]

\addplot[fill=highpp, bar width=0.2cm] coordinates {
    (7,\censusq)[7/11]
    (8,\lawq)[8/11]
    (9,\needq)[9/11]
    (9,\skepticismq)[9/11]
    (9,\validationq)[9/11]
    (8,\syntheticq)[8/11]
};

\addplot[fill=lowpp, bar width=0.2cm] coordinates {
    (0,\censusq)[0/6]
    (4,\lawq)[4/6]
    (4,\needq)[4/6]
    (5,\skepticismq)[5/6]
    (5,\validationq)[5/6]
    (2,\syntheticq)[2/6]
};

\addplot[fill=overall, bar width=0.2cm] coordinates {
    (7,\censusq)[7/17]
    (12,\lawq)[12/17]
    (13,\needq)[13/17]
    (14,\skepticismq)[14/17]
    (14,\validationq)[14/17]
    (10,\syntheticq)[10/17]
};

\legend{High PP, Low PP, Overall}

\end{axis}
\end{tikzpicture}
\caption{Absolute count of participants discussing general topics, out of a total of 17 participants. Presented overall, as well as sorted by participants with above average privacy priors (``High PP'', $n=11$) and below average privacy priors (``Low PP'', $n=6$). Most participants expressed a need or desire for differentially private synthetic data and carefully thought about the legal ramifications; many also expressed skepticism and a desire for real-data validation. However, only some participants had concrete experience working with synthetic data and \textit{none} of the participants with a low PP score mentioned the U.S. Census use case.}
\label{fig:percent}
\end{figure*}
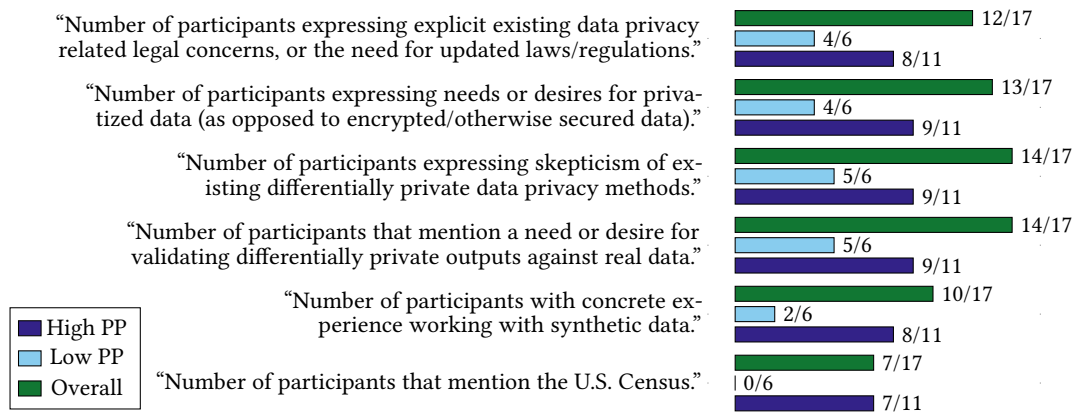


\subsubsection*{Communicating Data Privacy to Policy Makers} Citizens may trust authorities to follow best practices, but technical expertise is often lacking. \threepp noted this challenge with differential privacy for Census data, describing how his colleague ``made hundreds of presentations ... to the National Conference and State Legislatures ... [which were] not always ... well received. But ... even if we disagreed ... we agreed that we had been sufficiently transparent.''

\tenpp offers guidance to policymakers: ``My boss and I [are in a] partnership with [state education agencies], and are putting together a ... list for private synthetic data, multiparty encryption, etc. ... [It] says 'Here's what it's good for, here's what it's not typically good for.''' For example, ``for synthetic data, it's not great for longitudinal data, for geo-spatial data, [that's] pretty difficult.'' This leads to progress, as \tenonly realized: ``Usually people are ... happy ... They can [ask] more complicated [questions] when they [understand] the nuances ... [Questions like], 'What is a trusted execution environment compared to a secure enclave?' ... They'll say, 'Oh, I didn't realize I was technically using [Privacy Enhancing Technologies].' So that's really cool to point out.''

\subsection{``It's a chicken and the egg problem.'' - Findings Related to \RQone}

Below we discuss themes and highlight representative quotations relating to the question(s) that \RQone poses: \textit{``In what contexts do \dataexperts create, share, and use differentially private synthetic data in their work? What benefits do \dataexperts perceive in its potential use?''} 

 \begin{table*}[h!]
   \centering
 \caption{\RQone key takeaways: Perceived benefits of data privacy and private synthetic data.}
  \begin{tabular}{|p{2.5cm}|p{12cm}|}
    \hline
    \rowcolor[gray]{0.9} \textbf{Perceived Benefit} & \textbf{Explanation} \\
    \hline
    Enhancing \newline Research Collaboration and \newline Generalizability & \fifteenonly, \fourteenonly and \sevenonly all discussed how the ability to share data more freely, with strong privacy guarantees, could be a boon for research and open up possibilities for underresourced labs. As \fifteenonly put it, \textit{``Right now, you do your tiny study in your tiny lab, and you find something, and then, it never gets [to] the next step ... people [stop] earlier because we cannot share the data.''} while \fourteenonly noted \textit{``a democratizing benefit if you don't have an advanced kind of access.''}  \\
    \hline
    \rowcolor[gray]{0.9} Testing and \newline Tinkering & \seventeenonly, \tenonly, \sevenonly all mentioned excitement about the prospect of using differentially private synthetic data. As \sevenonly put it, \textit{``Most researchers don't wanna run that risk of exposure, [but] they want real looking data to test algorithms.''} \tenonly recalled \textit{``hear[ing] a lot of `We want synthetic data for testing'''} \\ 
    \hline
    Fulfilling \newline Obligations \newline for Data Protection & \sixteenonly and \eightonly discuss how an unintended leakage of personal information can lead to broken contractual obligations and significant legal consequences, and that private synthetic data can help to avoid legal pitfalls. \eightonly notes the biggest concern in enterprise applications is ``an unintended leakage of personal information ... breaking contractual obligations.'' \\
    \hline
    \rowcolor[gray]{0.9} Moral Obligations & \tenonly, \sevenonly and \oneonly discussed directly the moral and ethical obligations to protect individuals' data, and many other participants implied as much. As \tenonly put it, they feel obligated to protect individuals, \textit{``from the beginning [of collection] until the release at the public-end.''} \\
    \hline
    Reproducibility & \fifteenonly and \fiveonly both brought up a major challenge in the sciences: reproducing published results, and viewed private synthetic data as a potential solution. As \fifteenonly suggested, \textit{``the use of publishing data, together with your manuscript and the code to make it reproducible ... this could be a really important use [of private synthetic data] ... more information [given] to the next scientist increases the chance that [my research] can be deployed in the real world ... [that is] super super important.''} \\
    \hline
  \end{tabular}
  \label{tab:rq2}
 \end{table*}

\subsubsection*{Responses to Medical Tabular Data Prompt}
As discussed in Section~\ref{sec:methodology}, participants were shown fake medical tabular data and asked if they considered it sensitive. \textit{All participants acknowledged the sensitivity of medical data.} \fifteenpp emphasized that ``Data privacy [is]... super important in [the medical] field... because people are talking about very difficult situations that could impact other areas [of] their lives.'' Concerns included re-identification and rising insurance premiums. \tenpp noted that with additional data like a zip code, an insurance company could ``find a Veena within that zip code... and charge her more.'' \thirteenpp worried about ``someone using this information to... overcharge for health insurance.'' Even \elevenpp, despite skepticism about re-identification ease, acknowledged that ``if you have high blood pressure, your insurance premium could go up... [sometimes] participation in some study could harm you.''

\subsubsection*{The Desire(?) for Differentially Private Data}
A tension exists between the promise of privacy and the practical implications of sensitive data use. \tenpp voices this dilemma, acknowledging the comfort derived from privacy: ``privacy is great... [And] maybe you're not okay with [your data being used for] personalized ads... So having private data... gives me a little bit of... happiness,'' but also notes that ``a life-saving drug that helps me stay alive is better. So, [let's say] to have privacy for one million people, maybe [the privatization procedure] means [some life-saving drug] is worse, and like 3 people die that wouldn't have died otherwise. Is that worth it? Is there a trade-off?''

One response is to limit the use of differentially private data to less critical applications. \tenpp highlights that they have ``heard a lot of 'We want synthetic data for testing,' whether it's an agency that wants to test their own tools or products... or some external vendor [who] wants to see what the rough data looks like.'' Synthetic data for initial exploration, product testing, or low-risk applications helps limit potential privacy harms.

Yet, the potential to ``unlock'' data access through privacy methodologies is alluring and can be mission-critical. \seventeenpp discusses how ``Privacy concerns are brought in immediately, even in the ideation stage... you might be thinking like, `Oh, we could do this... with this data set.' And the first thing [a lawyer] says is `Well, there's no way we're gonna be able to get access'... often there's such a conservative approach that researchers might cancel certain ideas off the bat... because of an issue with getting the data.''

The ambiguity between data privacy \textit{hurting} utility versus \textit{helping} by unlocking otherwise unpursued avenues of research is encapsulated by \fourpp: ``Here's the hard part: it's a chicken and the egg problem. Nobody will publish with differentially private synthetic data if others don't.''

\subsubsection*{Motivation and Guidance}
Despite the challenges and the ``chicken and the egg'' problem, there is a desire among researchers to conduct studies using sensitive data. As \sevenpp explains: ``Most researchers don't wanna run that risk of exposure, [but] they want data to test algorithms.'' Using sensitive data comes with risks, but rewards can balance out. \twopp offers a balanced perspective, acknowledging the difficulty of finding a definitive answer (``I'm not sure we're ever gonna come up with a convincing answer'') while underscoring the importance of using this data in modern discussions of risk and compliance. He cites intergenerational wealth mobility data, where ``there's been a lot of uproar,'' but ``they have given us a much better understanding of [socio-economic] mobility... [allowing] us to start thinking about ways to address these concerns and to come up with policies that are effective,'' which wouldn't have been possible without sensitive data access. \sevenonly offers an unexpected privacy concern: ``We have to keep private the geolocation of underwater tree stumps... [to protect against] antique furniture makers,'' illustrating the diversity of data types requiring protection and unforeseen consequences of exposure.

With ``State and local agencies asking for guidance,'' \tenpp praises a report ``the White House just released... on privacy-preserving data sharing analytics... [which is] the closest they got to saying something about PETs.'' Government guidance is paramount as people deal with ubiquitous data collection. As \ninepp notes, ``[Lots of] data is collected by smart grids [which] can [provide] a lot of sensitive information about us... [companies] can tell when we wake up, or when we go to office, when we come back, when we do our laundry.''

\subsubsection*{Draws of Differentially Private Synthetic Data} Some \dataexperts were especially optimistic about the benefits of adopting differentially private synthetic data, responding to increased awareness of data privacy concerns. Participants mentioned how institutions fear privacy breaches in a data-saturated world. \seventeenpp notes, ``Anytime that there's a large privacy infringement, this is a source of massive loss of user trust... Anytime you don't follow those regulations, it could cost you billions of dollars.'' It's worth noting that re-centering the data subjects while articulating existing risks in DP governance is discussed at length in Seeman~\etal~\cite{seeman2024between}.

Most participants considered existing data privacy practices insufficient. \onepp recounted the New York City's Department of Education stating in a hearing that upon learning of a data breach they ``would just break [their] contract with that vendor.'' \oneonly remained hopeful that data privacy tools could prevent vendors from having direct data access. \fifteenpp was excited about more ``easily sharing [data] with colleagues who maybe [can] merge it together with their data set... So we can test the generalizability of [our results].'' Echoing concerns about reproducibility in science~\cite{fanelli2018science}, \sixpp discusses how she ``tells [her coworkers] that we want to implement this particular [differentially private synthetic data approach] in the code... in case of a data breach, we would be safe because we wouldn't have real data just synthetic data.''
\subsection{``My fear is that we could draw the wrong conclusions'' - Findings Related to \RQtwo}  
Below we select representative quotations and discuss themes related to the question(s) that \RQtwo poses: \textit{``What concerns do \dataexperts have in using differentially private synthetic data in place of original data?  What limitations do they experience in practice?  What suggestions do they have to make differentially private synthetic data useful to them?''}

\begin{table*}
  \centering
 \caption{\RQtwo key takeaways: Concerns related to differential privacy and data privacy at large.}
  \begin{tabular}{|p{3cm}|p{11cm}|}
    \hline
    \rowcolor[gray]{0.9} \textbf{Concern} & \textbf{Explanation} \\
    \hline
    Reliability in Sensitive Domains & The main concern raised by participants: accuracy and reliability of predictions made using private synthetic data. \sixteenonly worries about erroneous downstream predictions in critical applications like healthcare. \fifteenonly, \fourteenonly and \twelveonly shared similar concerns about per quintile distributions, outliers, and mediation analyses/longitudinal data respectively. \\
    \hline
    \rowcolor[gray]{0.9} Impact on Vulnerable or Minority Groups & Participants worried about impacts on minority groups and individuals in vulnerable situations. \thirteenonly wondered ``\textit{Eventually the model will be applied to ... minority groups or outliers, like, how does that impact them?}'' \nineonly points out that synthetic data may not accurately represent outliers, and ``\textit{if it's in the critical scenario, for example if you're testing a new drug, then those outliers are important.}'' \\
    \hline
    Misuse and Bypassing Regulation & \seventeenonly, \sixteenonly, \sixonly expressed fears that private synthetic data and differential privacy are convenient tools that can ``rubber-stamp'' access sensitive data, even if they aren't particularly careful in their application of the technology. \seventeenonly: \textit{``We do use a little bit of synthetic data in a not very rigorous way. My sense is that [other] people use [synthetic] data a lot [with] differential privacy. Methods to ... bypass regulations? Haha.''} \\
    \hline
    \rowcolor[gray]{0.9} Ambiguity in Laws & 
    \sixteenonly, \sixonly, \oneonly and \fouronly note that existing legal ambiguities present challenges in effective data privacy implementations. \sixonly: \textit{``The laws are very vague, they do not actually tell us what [a proper] implementation looks like ... ''}  \\
    \hline
    Resource Constraints and Practical Feasibility & \twelveonly, \tenonly, and \sevenonly raised concerns over the practical feasibility of implementing private synthetic data solutions, considering resource constraints. \tenonly: \textit{``A big concern we get a lot is staff time and financial cost ... .``How much will this cost us? ... How do we maintain it? How will my staff have time for this?'''} \\
    \hline
    \rowcolor[gray]{0.9} Preserving Statistical Properties & \twoonly, \fouronly and \fiveonly worry about preserving statistical properties (e.g., covariances) in synthetic data, especially in economic and social science research. \twoonly offers that \textit{``It's almost impossible to be confident [in covariances] with current methods for synthetic data ... this can lead to biased estimates.''} \fouronly notes that it may be important to have \textit{``a very complex hierarchical, nonlinear distribution [to] avoid having things like pregnant men aged 75.''} \\
    \hline
  \end{tabular}
  \label{tab:concerns}
\end{table*}

\subsubsection*{Practical Limitations} 
Though data privacy tools like differential privacy are already operational in government and industry, there are still limitations in practical execution for many companies and researchers dealing with sensitive data. In industry, for example, data privacy can be an afterthought. \sixteenpp recalls that at his startup, data privacy issues ``were mostly handled by one person we really trusted with security. At the time, it seemed... reasonable. Looking back, maybe not.'' \sixpp notes that legal teams ``may not understand a particular implementation of a differentially private synthetic data [system]... it might be a much longer timeline to implement... [even if]... [it] might [seem] like... a quick fix.''

Some participants, despite their best attempts, struggled with existing differentially private synthetic data tools. \elevenpp tried to use ``differentially private synthetic data'' but ``it rarely worked out... I couldn't figure out how to [make it work] for time series synthetic data research,'' a known limitation \cite{frigerio2019differentially}. \eightonly discusses how at his large tech company, ``The only way to look at the [sensitive] data... is to use a differentially private synthetic data generator,'' but adds, ``you would [still need] to go through the privacy review process, and argue that whatever transformation [you] did... [is accurate] and... that [the privacy] is sufficient.''

Review and validation processes can be difficult to manage and maintain, as \fourpp and \twopp can attest. \fourpp notes that despite running ``a validation server\footnote{A validation server is a 'secure method for researchers to submit statistical programs to run on a subset of the confidential administrative data;' recent versions incorporate differential privacy~\cite{burman2019safely, barrientos2021differentially}} for 10 years,'' he knows ``the challenge that poses.'' He says, ``I don't think validation servers will solve the problem'' even if researchers ``switch to publishing their data in a differentially private way.'' Still, \twoonly argues that ``Until we develop some validation methodology... I don't know how much differential privacy we're going to have,'' acknowledging the desire to validate results on real data, adding ``We're gonna have to figure out better ways, and more automated ways to [streamline] this process.''

\subsubsection*{Suggestions to Make Private Synthetic Data Actually Useful for \DataExperts}

We heard from nearly every participant that ensuring \textit{consistency} between the private synthetic data and the original data is the main driver of its usefulness.  As \seventeenpp put it, ``We [need to] see that the outcomes of our analyses would be the same across real and fake data sets.'' How can privacy researchers facilitate this confidence? \fourteenpp discusses facilitating researchers ``spot-checking'' synthetic data to ensure it matches domain knowledge. For example, ``you see... demographers doing this with the census data. They'll say `Oh, look at this weird [count] record for a lighthouse [of] negative one. People live here, that doesn't make any sense.''' \fivepp, a psychologist, is ``usually interested in individual differences [between] multiple variables.'' She wants proof that the synthetic data ``[maintains consistency] within the larger joint distributions... I would... check this with some cross tabulations... and scatterplots...'' As with many participants, \fiveonly ``would like to see the differences between what I get on the synthetic data and what I get on the real data.'' \four suggests we shouldn't be afraid to ``force people to publish the data if they can... so I can rerun your analysis... so I can assess whether... your analysis is robust.'' More available data and analysis, he argues, means we can assess the effectiveness of privatization schemes.

\subsubsection*{The 2020 U.S. Census}
\label{sec:census}
The 2020 U.S. Census has become the exemplar of multifaceted challenges when balancing operational efficiency, data privacy, and data quality while adhering to legal and constitutional mandates. However, the Census is an anomalously constrained use case, subject to intense scrutiny \cite{kenny2021use}. \threepp notes the conflict in meeting the constitutional requirement to take the census every ten years while satisfying the statutory requirement (Title 13, U.S. Code) not to publish any private information that identifies an individual or business.

When handling the ``trillions'' of statistics the Census must publish, one must accept ``the trade-offs between privacy, protection and data usability'' and ``prioritize certain use cases'' according to \threepp. The 2020 Census's use of DP was motivated in part by redistricting under Section 5 of the Voting Rights Act. \threeonly noted that policymakers must ``specify voting districts of equal population for congressional districts... so you want accuracy on the statistics used to enforce the Voting Rights Act... [this is] a use case that is extremely difficult to satisfy with modern [privacy] techniques.'' \fourpp suggested that some ``[differentially private methods] can't be scaled because you can't know... all the possible structural constraints... and you get the conundrum that the 2020 Census had, where... you're going to [fudge some stuff] because the [privatized] quality just isn't good enough.''

\subsubsection*{New Challenges from LLMs and Gen. AI}
Recent advances in generative AI were top of mind for some participants. Few (if any) legal precedents exist addressing the open question of whether large language model (LLM) owners are responsible for disclosing the use of and remunerating the owners of training data.  Lack of understanding in the way that training data is encoded in model weights has slowed legal challenges. However, \sixteenpp comments on ``the current lawsuit [between] Open AI and the New York Times,'' explaining that ``[there are] different forms of attacks, like prompt injections, ... [that allow you] to actually regurgitate a lot of information that [the LLM] was trained upon'' and positing that `` ... it's the first lawsuit I've seen [related to LLM training data].'' None of the prior studies listed in Table~\ref{tab:rel_work} report perspectives on differential privacy and recent advances in generative AI and LLMs; as far as we are aware, no such focused study has been conducted, and our study motivates this future work.

There is evidence that LLMs can be manipulated to divulge sensitive data, but it is unclear whether there are reasonable regulatory steps to mitigate potential harms. \sixteenonly: ``One of the issues [with regulating] LLMs is the amount of knowledge they can encode in their parametric memory ... [even if] there was some sort of legislation capping the number of model parameters ... you can distill a good amount of memory ... into smaller models ... so I think legislation on the data side makes more sense.'' Training these large language models effectively requires so much data that the aggressive data collection techniques risk including highly sensitive information. At small companies or startups, best practices may be ignored to reduce time to deploy. \sixteenonly recounted his experience ``using LLMs [to] create synthetic data based upon (type omitted) conversations ... [There was a lot of] financially incriminating data ... extremely private data ... We should've definitely [privatized] them.'' 
Scrubbing qualitative data presents a major challenge in collecting textual corpuses, as \twelvepp observes, ``sometimes [participants] will write answers to things that could identify them [or] they'll put their names or other people's names in [the data] by accident.''

Unfortunately, rigorous formal privacy constructs like differential privacy, which was designed to cleanly consider sensitive tabular datasets, do not adequately address large language models that incorporate billions of examples of textual data from potentially overlapping authors. \fourteenpp unpacks this limitation: ``So there's membership inference, right? You can figure out that someone's data was included in the model ... But if its a large language model, I kinda know that I'm in there somewhere [because] I'm on the Internet ... so I feel [that this scenario], the really well defined tabular one, has limits ... it's still very unclear [how LLMs] ... constitute a privacy leak or a leak of personal information.'' In response, \fourteenonly recommends considering data privacy for LLMs as ``a harms reduction approach'' as an initial attempt for safeguarding people's data used to train these large models.
\section{Discussion}
\label{sec:discussion}
Responses from participants addressing \RQone and \RQtwo yielded a wide range of findings, detailed extensively in the previous section. Below, we contextualize these findings by aligning them with prior work and integrating them into a broader narrative on DP adoption. We further discuss how our findings and analyses inform targeted \textbf{recommendations}, suggestions for norms and tools designed to meet the needs of the \dataexperts we interviewed.

\subsection{Privacy Solutions Must Be Context-Aware and Human-Centered.} 

Consistent with prior work on data privacy and DP frameworks~\cite{sarathy2023don, DBLP:journals/popets/GarridoLMS23}, our participants focused on the nuances of implementing robust privacy mechanisms, specifically for private synthetic data. Participants discussed challenges like domain-specific privacy requirements and communication barriers, reinforcing findings from other qualitative studies on privacy (Kyi et al.~\cite{kyi2024doesn} and Flanagan et al.~\cite{flanaganredesigning}). Additionally, our findings highlight contextual and human-centered design as a key theme when considering DP synthetic data. For instance, participants emphasized that a successfully private system must be \textit{flexible}. Privacy expectations that might be appropriate in healthcare may not be appropriate for social media. Thus, presenting universal toolings or use-cases for DP synthetic data as ``one size fits all'' may not effectively serve diverse contexts of use.

To help bridge the gap between different privacy expectations raised by participants, we recommend attaching context to use: \textit{(Recommendation 1) DP researchers and synthetic data offerings should commit to providing \textbf{evidence of validation} in at least \textbf{one partner-vetted use case}}. In other words, when presenting practical or open-source tools for differentially private synthetic data, it should be the norm to propose a real-world use (and this use should be reviewed by an application-oriented partner). This is in contrast to existing sanitized public benchmarks (the ``one size fits all approach''), which are divorced from their application specific context and cleansed to emphasize expected correlations at the expense of unexpected challenges (missing data, rare classes, spurious correlations).

\subsection{DP Synthetic Data Can Unlock New Research Opportunities, But Trust and Consistency Are Key}
When asked about differentially private synthetic data (\RQone), participants acknowledged both its promise and its practical limitations, but were generally aware of the growing availability of private synthetic data tools~\cite{horvitz2015data, jain2016big, diffprivlib, Shoemate_OpenDP_Library, zhang2017privbayes, rosenblatt2020differentially, mckenna2021winning, liu2021iterative, mckenna2022aim}. As \eightonly mentioned, ``the only way to look at sensitive data as a data scientist is to use differentially private synthetic data,'' but concerns about consistency between the synthetic and real datasets persisted. Several participants voiced skepticism about the practical benefits of DP in mission-critical domains, such as healthcare or drug discovery, where accuracy is paramount. The tension between the ideal of privacy and the reality of reduced utility was palpable as participants frequently weighed trade-offs in their responses. As prior work by van Hoorn et al.~\cite{van2024acceptance} noted, privacy tensions arise when sensitive data is necessary for critical applications, such as drug discovery (mentioned by multiple participants in our study), which relies heavily on accurate data. Participants generally supported the idea that privacy tools can unlock new avenues for research by enabling safer data sharing, even if some were skeptical of its practical benefits. Approaches to dealing with this tension -- between privacy and utility -- reflect an ongoing debate in privacy-preserving technologies~\cite{sarathy2023don}.

In the narrower context of DP synthetic data, participants specifically pointed to \textit{trust} as the critical factor in determining utility. Several expressed the need for validation methods that can demonstrate consistency between synthetic and original data, a key concern also noted by subjects in Garrido et al.~\cite{DBLP:journals/popets/GarridoLMS23}. Without reliable methods for comparing the two, it remains difficult for \dataexperts to fully trust synthetic data for high-stakes applications. As \seventeenonly succinctly put it, ``we need to see that the outcomes of our analyses would be the same across real and fake datasets.'' But what constitutes ``the same'' across data contexts? We suggest that here, too, \textit{(Recommendation 1)} can help; \textit{partner vetted evidence of validation} would help build necessary trust. 

Participant responses suggested we must go further still when building trust in DP systems: \textit{evidence of validation} must be \textit{widely and transparently shared}. This leads to \textit{(Recommendation 2): organizations consuming differentially private data should agree on, and publish, their \textbf{standards of evidence}.} This goes beyond simply ``exposing your epsilons'' (as was proposed by Dwork et. al \cite{dwork2019differential}); comprehensive standards should advise on restrictions and best practices across \textit{all} stages of the data life-cycle.  

\subsection{The Communication Gap Around Differential Privacy Is (Still) a Major Barrier to Entry}  

Participants spoke extensively about commonly understood communication challenges around difficult to grasp concepts (such as the aforementioned $\epsilon$ parameter in DP), which was consistent with earlier research by Nanayakkara et al.~\cite{nanayakkara2023chances}. Our interviews reinforced the existence of a gap between theoretical models of privacy and their real-world applications; participants went so far as to suggest an \textit{epistemological} barrier for most \dataexperts. As \tenonly reflected, ``even if you say it's mathematically private, people don’t always understand what that means in practice.'' To this end, participants expressed a need for better educational resources and communication tools to demystify DP, a sentiment shared by subjects of prior studies~\cite{DBLP:conf/chi/BullekGMP17}.

However, our findings suggest that we must go beyond a focus on communicating DP details (like $\epsilon$). The major communication barrier is that many \dataexperts perceive a \textit{lack of conceptual consensus} within their community, not that some \dataexperts lack an understanding of some singular DP concept. In other words, communicating about singular DP concepts can only be effective if \dataexperts have clear community standards that motivate learning. 

\textit{(Recommendation 2)} can also help here: by aligning community standards and norms with the shared training of an organization's members, the community as a whole can reach a \textit{conceptual} privacy consensus, thus helping to motivate community members to learn more granular DP concepts where applicable. For example, statisticians or economists may use precise characterization of the error introduced by the privacy mechanism as a standard of evidence, while empiricists or social scientists may tailor application-specific benchmarks for their standard of evidence. Each group will lean on different DP concepts and thus require different learnings.

A number of our participants were excited about visual tools and intuitive explanations, which could  be instrumental components in communicating DP standards of evidence. Recent work expands these available resources, and we encourage further research in this vein \cite{nanayakkara2023chances}. Without such intuitive visualization tools, mathematically sound privacy frameworks risk being misunderstood and underutilized.

\subsection{Epistemological Gaps are Obstacles to Practical Adoption}

Participants highlighted several gaps in current DP tooling that hinder broader adoption (\RQtwo), particularly in the public sector. In general, this finding aligns with Garrido et al.~\cite{DBLP:journals/popets/GarridoLMS23}, who found that organizations often face practical gaps in DP tooling, limiting their ability to effectively use these technologies. Participants in our study communicated a more nuanced, expanded message. They highlighted \textit{epistemological} gaps in understanding the \textit{usefulness} of private synthetic data. These gaps are rooted in a \textit{fundamental uncertainty} about real-world applications for DP synthetic data. This overwhelming lack of clarity around the capabilities of DP synthetic data undermines deployment. \tenonly emphasized the dilemma this creates: ``privacy is great ... but the trade-off is always there: are you sacrificing too much utility for too little privacy?''

Multiple participants highlighted one potential, real world use case for existing DP synthetic data that could improve data access \textit{without} needing to directly contend with epistemological concerns of utility. We summarize this concretely as \textit{(Recommendation 3): Use DP synthetic data to create a tiered access model for access to sensitive data, where promising results on high-privacy but low-fidelity synthetic data can be used to apply for greater access to higher-risk differentially private data, or to the original data itself.}  We call this a \textit{``driver's license''} model, as a potential data user could be given a ``permit'' to perform ad hoc and exploratory research before being given a ``license'' to work with the real, potentially sensitive data.

Despite these suggestions for potential DP synthetic data deployment models, participants generally highlighted how trade-offs between privacy and utility lead to skepticism, some with a particular focus on the 2020 U.S. Census (as we discussed in Section~\ref{sec:census}). As boyd and Sarathy~\cite{boyd_2022} detail, the Census Bureau's use of DP synthetic data for census data was met with a severe backlash. Participants in our study with an extensive familiarity with the Census Bureau's use of DP found that it exemplified the delicate balance between protecting privacy and maintaining data quality. Conversely, participants who were less familiar felt the Census use case as an example, felt it did not fully capture the diversity of challenges faced by other sectors, and suggested that considerations for census data are not a universally applicable model for DP governance. As a whole, we found a clear disconnect between the census use case and participants with the least data privacy experience, who \textit{did not mention the Census at all} (Figure~\ref{fig:percent}). This leads us to believe that privacy experts have overfit their thinking on DP deployment, specifically for synthetic data, to the Census use case.

\subsection{Limitations}\label{sec:limsec}
Despite the comprehensive insights gained from our interviews, we'd like to highlight several limitations of our study. We reached saturation with respect to our research questions with our sample of 17 participants; that said, these research questions may not have \textit{fully} captured the broad spectrum of perspectives and experiences that could exist on differentially private synthetic data. Additionally, our findings are inherently shaped by the specific contexts and disciplines of the interviewees, which may limit the generalizability of our conclusions. Future work should aim to expand our list of research questions, alongside the participant pool, to ensure a more holistic understanding of the challenges and opportunities in deploying differentially private synthetic data.

\section{Conclusion and Future Research Directions}
\label{sec:conc}

Our interviews prompted us to consider remaining open questions 
around differentially private synthetic data. We advocate for  
research that improves systems-oriented infrastructure for differentially private data sharing, widens our understanding of best methods for communicating and teaching complex topics in data privacy to \dataexperts and policy makers, and solidifies practical threat models.
In each area, we emphasize a stakeholder-first approach that centers the specific applications rather than the general algorithms.

Our findings and analysis reveal that although \dataexperts conceptualize data privacy differently and vary in their understanding of differential privacy, they agree on its importance. That said, our participants raised concerns about the generalizability of results and the potential harm of misinforming individuals or underrepresented groups. These concerns appear to prevent the uptake of differentially private synthetic data in practice; respondents considered it more as a last resort despite their ability to clearly articulate the potential benefit of broadening research access to sensitive data.

Based on our findings, we proposed three key recommendations. The first was \textbf{Evidence of Validation}: all participants agreed on the necessity of validating synthetic data against real data and outcomes of interest, and private mechanism design should incorporate this belief. The second was \textbf{Standards of Evidence}: those who use differentially private synthetic data should publicly establish their standards for evaluating its efficacy. These standards will inevitably vary, requiring precise error characterization and favoring application-specific benchmarks. The third was a \textbf{Tiered Access Model}: this tiered access approach to sensitive data would allow researchers, particularly those in smaller labs or working individually, to initially access high-privacy, low-fidelity data. Successful and responsible use of this data could then lead to access to higher-risk differentially private synthetic data or even the original datasets. Our recommendations acknowledge the iterative nature of research and provide a practical framework for a path forward managing privacy risks in a variety of research and industry settings.


\bibliographystyle{ACM-Reference-Format}
\bibliography{software}

\newpage
\appendix
\section{Ethical Considerations, Researcher Positionality, and Potential Adverse Impacts}

Responsible research mandates reflection on positionality and potential adverse impacts of our work.
A significant risk inherent in an interview study like ours is the potential for the misinterpretation or decontextualization of quotations. We emphasize the importance of including the specific context and prompting that elicited responses from our participants in any future 
discussions of our work. 

Our positionality is that of researchers developing data privacy tools and evaluation techniques, studying engagement with these tools, and considering their broader impacts on society. Our bias includes an interest in technological solutions and an interest in the epistemological foundations. We strove to mitigate the bias by including a wide variety of direct quotes, with thorough context to aid in interpretation.

We received IRB approval at our institution to conduct this study, and took care to remove identifiable information from all data. However, to accurately position each participant and provide context for their perspectives, we included background information that could be considered identifiable. We believe this inclusion was necessary to provide evidence of our conclusions, though we acknowledge a small but non-zero risk of re-identification.  
All participants were informed about the potential inclusion of their background information in the final paper and provided consent.

\section{Interview Questions}
\label{app:questions}
Here we provide the outline of questions we referenced during our semi-structured interviews. These were used by the interviewer to stay on track, but the interviewer deviated from these questions where appropriate. Question wordings were slightly modified upon each asking, so as to preserve the conversational quality of each interview. One of the authors conducted all interviews for consistency. 

\subsubsection{Opener} ``While I have some questions prepared, please know that this conversation is meant to be fluid and dynamic. If you would like to share information or experiences that diverge from the specific questions I ask, please do so. Our goal is to ensure a comprehensive understanding of your perspective, so please feel free to express yourself openly and honestly.
\textit{Who you are:}''

\subsubsection{About} I’m going to begin with 4 basic questions just to get a sense of who you are and what you do. \textit{To start:}

\textbf{Q1:} Between, ``Public (Government, etc.)'', ``Private,'' or ``Academic,'' how would you describe the sector of your work? \textit{Feel free to list a mix of sectors or introduce your own.}

\textbf{Q2:} We are hoping to provide a sense of the gender, race, ethnicity and educational backgrounds in our study, in order to demonstrate varied perspectives of participants. If you feel comfortable doing so, \textit{please share how you would self-identify.} Note that your exact answer won’t be shared, only aggregated to give an overall view of all participants.

\textbf{Q3:} \textit{Do you work on a specific application domain for your work? And if so, how would you characterize it?}

\subsubsection{Data Privacy}

\textbf{Q1:} \textit{How would you characterize the term ``data privacy'' in your own words OR what does the term ``data privacy'' mean to you?}

\textbf{Q2:} \textit{Do you encounter what you would consider to be sensitive data in your own work?}

\textbf{Q3:} \textit{In your own words, what qualifies as sensitive data?}

\textbf{Q4:} Not necessarily in your own work, but in general, \textit{are you familiar with any data breaches or privacy violations?}

\subsubsection{Slide prompting: Tabular Data and Terms}

\textbf{Q1:} \textit{Can you recall times where you could not access important data for your research due to privacy or other restrictions?}

\textbf{Medical Tabular Data Example} \textit{To you, does this data qualify as sensitive data? If so, what makes it sensitive?}

We'll now go through some slides that have terms from the field of data privacy on them. Please offer a definition in your own words of each term if you are able, or, if you are are unfamiliar with the term, don't worry I will ``pop'' up the definition for you.

\textbf{De-identified data} The first term is ``de-identified data'' -- \textit{how would you characterize this term in your own words? }

\textbf{K-anonymity?}

\textbf{Differential Privacy?}

\begin{figure*}[t]
\includegraphics[width=\textwidth]{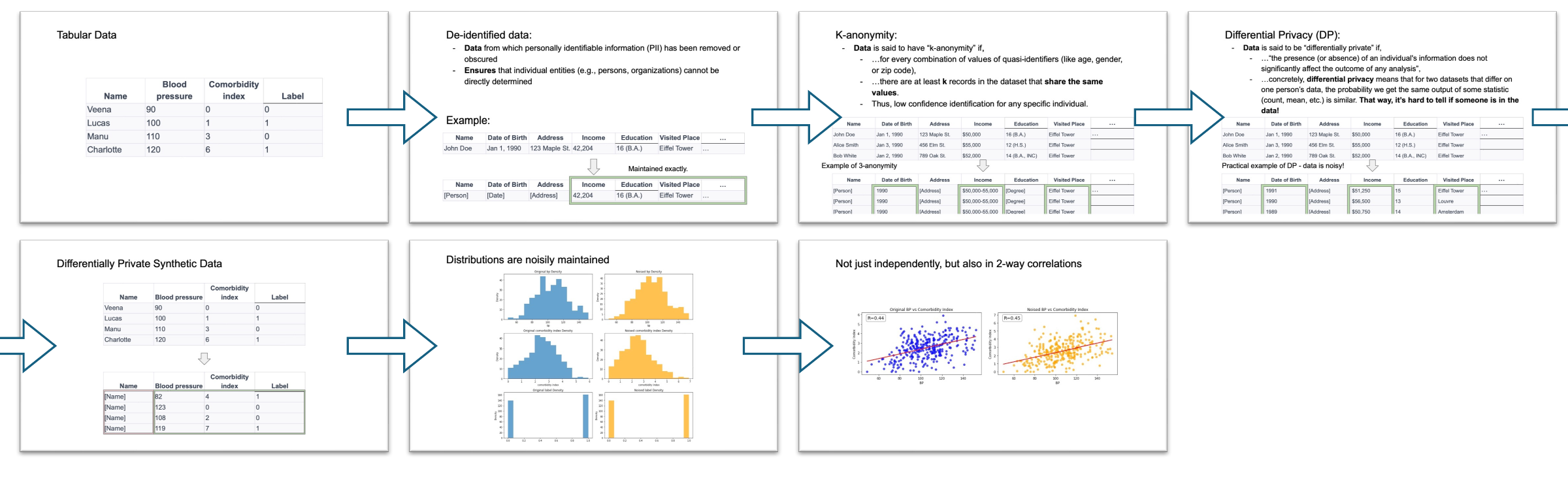}
\caption{Full slides used for participant prompting during interviews, abbreviated in Figure~\ref{fig:slides}}
\label{fig:full_slides}
\end{figure*}

\subsubsection{Slide prompting: Private Synthetic Data}
We're now going to go through a hypothetical example with differentially private synthetic data. But, before we do that...

\textbf{Q0:} \textit{How would you characterize what ``synthetic data'' is in your own words?}

\textbf{Q1:} \textit{Have you ever worked with synthetic data, or, if not, do your peers work with synthetic data? How have they used it?}

\textit{Show participant private synthetic medical data example} Here we have some differentially private synthetic data. We have achieved this by adding ``noise'' to the distribution of the original data. So, on their face, the densities of each variable look similar. Also, some correlations in the data were preserved, like this one between blood pressure and co-morbidity.

\textbf{Q2:} \textit{Are the comparisons and figures I just showed you convincing? If so, what was convincing? If not, what would be a more convincing set of metrics, plots or comparisons?}

\textbf{Q3:} \textit{Would you ever use this type of private synthetic data in your own work?}

\textbf{Q4:} \textit{In a hypothetical scenario where you are a medical researcher and this is the actual data you have access to, would you publish your results? If not, what would prevent you from doing so?}

\newpage
\section{Heirarchical Codebook}
\label{app:codebook}

\begin{table}[htbp]
\centering
\begin{tabular}{p{2cm}p{3cm}p{8cm}}
\toprule
\textbf{Code} & \textbf{Description} & \textbf{Example Coded Quotation} \\ \midrule
Access & Availability or ability to use data, resources, or information & \twelveonly: `People are not clear on ... how to keep data private or how to share data or [even] where can data be stored?' \\ \midrule
Data management & Organization, storage, and maintenance of data & \twelveonly: `...there're still times where [I ask], Can I have access to this? How should I store this? How can I send this?' \\ \midrule
Data privacy & Practices or methods to protect personal or sensitive data & \tenonly: `If statistics are released on a data set, those individuals [should be] protected ... from the beginning [of collection] until the release.' \\ \midrule
Data quality & Accuracy, completeness, and reliability of data & \nineonly: `Synthetic data may not accurately represent outliers, and ... in a critical scenario, for example if you're testing a new drug, then those outliers are important.' \\ \midrule
Data security & Measures to protect data from unauthorized access or corruption & \oneonly: `New York City's Department of Education [stated] ... if there was a data breach, they `would just break [their] contract with that vendor.'' \\ \midrule
Privacy (definition) & Specific definitions or frameworks for protecting data & \sevenonly: `In healthcare, I expect one level of privacy, and with social media, I expect a different level of privacy.' \\ \midrule
Synthetic data (use) & Application or analysis of synthetic datasets & \tenonly: `My boss and I [are] putting together a ... list for private synthetic data ... Here's what it's good for, here's what it's not typically good for.'
 \\ \bottomrule
\end{tabular}
\caption{Data Management and Privacy}
\end{table}

\begin{table}[htbp]
\centering
\begin{tabular}{p{2cm}p{3cm}p{8cm}}
\toprule
\textbf{Code} & \textbf{Description} & \textbf{Example Coded Quotation} \\ \midrule
Analysis & Examining data to draw conclusions & \seventeenonly: `We need to see that the outcomes of our analyses would be the same across real and fake data sets.' \\ \midrule
Correlation & Relationship between two or more variables & \fivepp: `I would probably want to check this with some cross tabulations ... and scatterplots that index by different facets ...' \\ \midrule
Statistic & Numerical measurement derived from data analysis & \fiveonly: `I would like to see the differences between what I get on the synthetic data and what I get on the real data.' \\ \midrule
Research methods & Specific approaches to gather and interpret data & \fourpp summarizes a typically chaotic research process: `I'm gonna download a bunch of variables that I think are relevant ... and then ... use a subset ... in my paper.' \\ \midrule
Uncertainty & Lack of certainty in data, findings, or analysis & \twopp: `I'm not sure we're ever gonna come up with a convincing answer [to data privacy issues].' \\ \midrule
Validity & Extent to which data or research methods accurately measure their intended objectives & \fouronly: `It's important to have a very complex hierarchical, nonlinear distribution [to] avoid having things like pregnant men aged 75.' \\ \midrule
Distribution & How data or values are spread across a dataset & \twoonly: `It's almost impossible to be confident [in covariances] with current methods for synthetic data ... this can lead to biased estimates.' \\ \bottomrule
\end{tabular}
\caption{Research and Statistical Methods}
\end{table}

\begin{table}[htbp]
\centering
\begin{tabular}{p{2cm}p{3cm}p{8cm}}
\toprule
\textbf{Code} & \textbf{Description} & \textbf{Example Coded Quotation} \\ \midrule
People & Individuals or populations involved in the study & \twelvepp notes that in her lab, `People are not clear on ... how to keep data private or how to share data ...' \\ \midrule
Population & Group of individuals or entities being studied & \thirteenonly: `Eventually the model will be applied to ... minority groups or outliers, like, how does that impact them?' \\ \midrule
Concerns & Worries or potential issues in a domain & \ninepp: 'It's so confusing, `How do you select the proper epsilon value? What does [it] actually mean?... Sometimes, it feels like it's just for mathematical purposes.' \\ \midrule
Context & Background or setting in which data is situated & \fourpp: `It really depends on the socio-economic, legal context of how the data was collected, how it is being disseminated, and what the intended use of it is.' \\ \midrule
Contextualization & Placing data or findings in the appropriate context & \twopp: `... intergenerational wealth mobility data [gives] us a much better understanding of [socio-economic] mobility, allowing us to start thinking about ways to address these concerns.' \\ \midrule
Legal and ethical considerations & Legal and moral aspects of research or data management & \tenpp: `... have heard a lot of 'We want synthetic data for testing ... for lower-risk applications, it helps limit potential privacy harms.' \\ \midrule
Social issues & Concerns related to society, often linked to data or research & \seventeenpp: `[In] some ... countries around the world, privacy concerns [are] pushed aside in certain areas for the public good.' \\ \midrule
Trust & Belief in the reliability or accuracy of data, systems, or research & \seventeenpp: `Anytime there's a large privacy infringement, this is a source of massive loss of user trust ...' \\ \bottomrule
\end{tabular}
\caption{Social and Organizational Context}
\end{table}

\begin{table}[htbp]
\centering
\begin{tabular}{p{2cm}p{3cm}p{8cm}}
\toprule
\textbf{Code} & \textbf{Description} & \textbf{Example Coded Quotation} \\ \midrule
Decision-making & Choosing between alternatives based on data or analysis & \fourpp: `Here's the hard part: it's a chicken and the egg problem. Nobody will publish with differentially private synthetic data if others don't.' \\ \midrule
Development & Growth, improvement, or creation in a professional or technical context & \sixpp: `[We want to implement this private synthetic data approach] in case of a data breach ... we would be safe because we wouldn't have real data, just synthetic data.' \\ \midrule
Efficiency & How resources (e.g., time, data) are used to achieve outcomes & \tenonly: `A big concern we get a lot is staff time and financial cost ... How much will this cost us? ... How do we maintain it?' \\ \midrule
Application & Use or implementation of data or methods in practice & \tenpp: `[We are] putting together a ... list for private synthetic data, multiparty encryption ... Here's what it's good for, here's what it's not typically good for.' \\ \bottomrule
\end{tabular}
\caption{Decision-Making and Development}
\end{table}

\begin{table}[htbp]
\centering
\begin{tabular}{p{2cm}p{3cm}p{8cm}}
\toprule
\textbf{Code} & \textbf{Description} & \textbf{Example Coded Quotation} \\ \midrule
Emotion & Emotional reactions associated with data or outcomes & \tenpp: `Privacy is great ... having private data gives me a little bit of happiness ... but [saving lives with the data] is better.' \\ \midrule
Identity & How individuals perceive themselves within a study or dataset & \fifteenpp: `[Medical] data privacy is super super important ... because people are talking about very difficult situations that could impact other areas [of] their lives.' \\ \midrule
Psychological factors & Mental or emotional considerations that influence behavior & \fifteenonly: `People are talking about very difficult situations that could impact other areas [of] their lives.' \\ \midrule
Agency & Capacity to act independently and make free choices & \sixteenonly: `...do we have any autonomy over where that data goes? ... Data privacy encapsulates all of that.' \\ \bottomrule
\end{tabular}
\caption{Emotion and Identity}
\end{table}

\begin{table}[htbp]
\centering
\begin{tabular}{p{2cm}p{3cm}p{8cm}}
\toprule
\textbf{Code} & \textbf{Description} & \textbf{Example Coded Quotation} \\ \midrule
Attack & Threats or vulnerabilities in data security or privacy contexts & \tenpp: `... an unintended leakage of personal information [breaking contractual obligations].' \\ \midrule
Risk & Potential negative outcomes or vulnerabilities in a system or dataset & \tenpp: `[If] the privatization procedure means [a life saving drug] is worse, and like 3 people die that wouldn't have died otherwise. Is that worth it? Is there a trade off?' \\ \bottomrule
\end{tabular}
\caption{Adversarial Concerns}
\end{table}

\begin{table}[htbp]
\centering
\begin{tabular}{p{2cm}p{3cm}p{8cm}}
\toprule
\textbf{Code} & \textbf{Description} & \textbf{Example Coded Quotation} \\ \midrule
Health & Medical or well-being aspects connected to data or research & \fifteenpp: `Data privacy [is] super super important in [the medical] field ... people are talking about very difficult situations that could impact other areas [of] their lives.' \\ \midrule
Academia & Individuals or activities related to higher education or research & \onepp discusses these challenges, `... if I'm running an academic research project I will have gone through an Ethics Review ... [but as a] consultant ... there isn't a formal data agreement ... I just have free access to some of their documents and resources.' \\ \midrule
Census & Systematic collection and recording of population data & \threepp: `... in using differential privacy for Census data, describing how his colleague `made hundreds of presentations ... to the National Conference and State Legislatures ... [these presentations were] not always ... well received.'' \\ \midrule
\end{tabular}
\caption{Other Codes}
\end{table}

\end{document}